\documentclass[prd,aps,showpacs]{revtex4}
\usepackage{amsmath}
\usepackage{pdfpages}
\usepackage{graphicx}   
\usepackage{array}
\newcommand{\du}{{d_\mathcal{U}}}
\newcommand{\U}{{\cal U}}

\begin{document}

\title{Analysis of $\mu-\tau$ conversion through $\mu N\to \tau X$ deep inelastic scattering induced
by unparticles}
\author{A. Bola\~nos}		
\affiliation{Facultad de
Ciencias F\'\i sico Matem\'aticas, Benem\'erita Universidad
Aut\'onoma de Puebla, Apartado Postal 1152, Puebla, Pue., M\'
exico}
\author{A. Fernandez}		
\affiliation{Facultad de
Ciencias F\'\i sico Matem\'aticas, Benem\'erita Universidad
Aut\'onoma de Puebla, Apartado Postal 1152, Puebla, Pue., M\'
exico}
\author{ A. Moyotl}
\affiliation{Facultad de
Ciencias F\'\i sico Matem\'aticas, Benem\'erita Universidad
Aut\'onoma de Puebla, Apartado Postal 1152, Puebla, Pue., M\'
exico}
\author{G. Tavares-Velasco}		
\email[E-mail:]{gtv@fcfm.buap.mx}
\affiliation{Facultad de
Ciencias F\'\i sico Matem\'aticas, Benem\'erita Universidad
Aut\'onoma de Puebla, Apartado Postal 1152, Puebla, Pue., M\'
exico}

\date{\today}
\begin{abstract}
A study of  $\mu-\tau$ conversion via  the deep inelastic scattering (DIS) process $\mu N\to \tau
X$, with $N$ a nucleon, is performed  taking into account
the effects from both spin-0 and spin-1 unparticles with
lepton flavor violating (LFV) couplings. This process has attracted attention in the past
as it may be at the reach of a future neutrino or muon factory. For the model
parameters, we use the
most recent constraints on the unparticle LFV couplings from the experimental limits on the muon
anomalous magnetic moment and the LFV decay $\tau\to 3\mu$, whereas for the
unparticle scale $\Lambda_\U$ and scale dimension $\du$ we use the bounds obtained from the
search for monojets
plus missing transverse energy
at the LHC. The $\mu N\to
\tau X$ cross section is analyzed when the target is a proton and it is found that the unparticle
effects
can be larger than the contribution from Higgs exchange in the minimal
supersymmetric
standard model (MSSM). We also analyze the behavior of the angular and energy distributions of the
emitted
tau lepton, which could be used  to disentangle among distinct new physics contributions. It is
found that, for a beam with an intensity of $10^{20}$ muons with  an
energy around 50 GeV on a $10^2$ gr/cm$^2$ mass target annually, there would be about $10^2-10^3$
$\mu
N\to \tau X$ events per year. The potential background is discussed briefly.

\end{abstract}
\pacs{11.30.Hv,13.85.Fb,13.85.Qk}
\maketitle

\section{Introduction}

Georgi has conjectured a hidden scale invariant
sector in the high-energy theory \cite{Georgi:2007ek,Georgi:2007si} that could interact
with the standard model (SM)
via scale invariant fields  associated with the so-called unparticles, a denomination due to the
fact that
scale invariant fields  with nontrivial anomalous dimension cannot be interpreted
in terms of particles.
Although the description of such a theory could be extremely complex, one can still study
its low-energy  effects through the
effective Lagrangian approach.
The  ingredients to describe the effective Lagrangian that parametrizes the unparticle
interactions with
the SM can be found in  \cite{Banks:1981nn}: the hidden
sector is a Banks and
Zaks $({\mathcal B}{\mathcal Z})$ sector and the associated fields are introduced through
renormalizable operators ${\mathcal O}_{{\mathcal B}{\mathcal Z}}$. The interaction of this sector
with
the SM fields occurs through the
exchange of heavy particles at a very high-energy scale $M_{\U}$. Below such a
scale, there emerge
nonrenormalizable couplings between the ${\mathcal B}{\mathcal Z}$ sector and the SM. As
scale
invariance emerges, dimensional transmutation proceeds via the renormalizable couplings of
the ${\mathcal B}{\mathcal Z}$
sector at an energy scale $\Lambda_{\U}$. An effective theory can
describe the interactions of the ${\mathcal B}{\mathcal
Z}$ and SM fields, which occur via unparticles.
The effective Lagrangian can be written as
\cite{Georgi:2007ek,Georgi:2007si}:

\begin{equation}
{\mathcal L}_{\U}=C_{{\mathcal O}_{\U}} \frac{\Lambda_{\U}^{d_{{\mathcal B}{\mathcal
Z}}-\du}}{M_{\U}^{d_{SM}+d_{{\mathcal B}{\mathcal Z}}-4}} {\mathcal
O}_{SM}{\mathcal O}_{\U}, \label{efflag}
\end{equation}
where $C_{{\mathcal O}_{\U}}$ stands for the coupling constant, whereas
the unparticle operator, ${\mathcal O}_{\U}$,  can be of fractional dimension $\du$.
The unparticle operators, which can be constructed out of the primary operators
${\mathcal O}_{{\mathcal B}{\mathcal Z}}$ and their transmutation, can be of scalar,
vector,
spinor, or tensor type. Unparticle propagators are constructed using unitary cuts and
the spectral decomposition
formula. By this means, the propagator of a spin-0 unparticle is found to be:
\begin{equation} \Delta_F(p^2) = \frac{A_{d_{\cal U}}}{2
\sin{(d_{\cal U} \pi)}} (-p^2-i \epsilon)^{d_{\cal U}-2},
\label{scapropagator}
\end{equation}
where the $A_{d_{\cal U}}$ function is introduced to
normalize the spectral density \cite{Cheung:2007ap} and is given as
follows:
\begin{equation} A_{d_{\cal U}}=\frac{16 \pi^2
\sqrt{\pi}}{(2 \pi)^{2 d_{\cal U}}} \frac{\Gamma(d_{\cal
U}+\frac{1}{2})}{\Gamma(d_{\cal U}-1) \Gamma(2 d_{\cal U})}.
\end{equation}
As for the propagator of a spin-1 unparticle, it is
\begin{eqnarray}
\Delta_F^{\mu\nu}(p^2)& = &\Delta_F(p^2)\left(
-g^{\mu\nu}+a \frac{p^\mu p^\nu}{p^2}\right).\label{vecpropagator}
\end{eqnarray}
The condition $a = 1$ is fulfilled when the
unparticle field is  transverse, namely,  $p_{\mu}
\Delta_F^{\mu\nu}(p^2) = 0$.


Unparticle phenomenology has been widely studied. For instance, peculiar  effects
arising from the interference between unparticle and SM contributions could show up  in
the Drell-Yan process at
hadronic colliders \cite{Cheung:2007zza,Mathews:2007hr,Kumar:2007af}.
The direct production of unparticles has also been studied in both leptonic
\cite{Cheung:2007ap}
and hadronic colliders \cite{Rizzo:2008fp}.
Not only the tree-level unparticle effects have been the focus of attention, but also
one-loop induced effects
\cite{Hektor:2008xu,Iltan:2007ve,Moyotl:2011yv,Ding:2008zza,Martinez:2008hm}:
the electron magnetic dipole moment via scalar and vector unparticles was first obtained
in
Refs. \cite{Luo:2007bq,Liao:2007bx}. This study was later extended for the lepton
magnetic moment due to  scalar \cite{Hektor:2008xu} and vector \cite{Moyotl:2011yv}
unparticles with lepton flavor
violating (LFV) couplings, whereas the lepton electric dipole moment via scalar
\cite{Iltan:2007ve} and vector \cite{Moyotl:2011yv}  unparticles was studied more
recently.
Other studies worth mentioning deal with the potential unparticle effects  on
CP violation \cite{Chen:2007vv,Huang:2007ax}, neutrino physics
\cite{Barranco:2009px}, etc.
Direct constraints on the
scale $\Lambda_{\U}$ and the dimension $\du$ have been extracted
from the LEP, Tevatron and LHC data. For instance, the $e^-e^+\to \gamma\U$ process was
studied to
explain $\gamma\bar\nu\nu$ production at LEP \cite{Cheung:2007ap}.  More recently the CMS
collaboration
 has imposed constraints on the unparticle parameters from the data of the search
for monojets plus large missing transverse energy at the LHC \cite{Chatrchyan:2011nd}.
Indirect
constraints have also been obtained from experimental data in cosmology,
astrophysics
\cite{Davoudiasl:2007jr,Hannestad:2007ys,Das:2007nu,Freitas:2007ip},  the
muon magnetic dipole moment (MDM), and LFV processes \cite{Hektor:2008xu, Moyotl:2011yv}.

As far as LFV is concerned, it is well known that any signal of this class of transitions would
be a clear evidence of new physics. Strong experimental constraints on
LFV muon decays have been placed that considerably disfavor this class of processes:
BR$(\mu \to e\gamma) < 2.4 \times 10^{-12}$ \cite{Adam:2011ch}, BR$(\mu\to 3e) < 1.0
\times
10^{-12}$ \cite{Bellgardt:1987du}, and BR$(\mu {\rm Ti} \to
e {\rm Ti}) < 3.6 \times 10^{-11}$ \cite{Kaulard:1998rb}.  On the other
hand, there are less stringent constraints on LFV tau  decays:  BR$(\tau\to e
\gamma)
\lesssim
10^{-8}$, BR$(\tau\to \mu
\gamma)\lesssim
10^{-8}$ \cite{:2009tk}, BR$(\tau\to 3e) < 3.6 \times
10^{-8}$ \cite{Aubert:2007pw}, and  BR$(\tau\to e^-e^+ \mu) < 3.7 \times
10^{-8}$ \cite{Miyazaki:2007zw}. Therefore, there is still a chance that $\mu-\tau$
transitions  may occur with a measurable rate. Such a possibility has been explored in
several SM extensions. In this work we are interested in the study of  $\mu-\tau$
conversion  via the deep inelastic scattering (DIS)
process $\mu   N \to \tau X$, where $N$ is a nucleon, in the context of  unparticle
physics. This process, which could be at the reach of a future neutrino or muon factory, has
attracted some attention during the past
\cite{Gninenko:2001id,Sher:2003vi,Kanemura:2004jt}. We will consider the
contributions from both spin-0 and spin-1 unparticles assuming the current bounds on the
unparticle scale and the LFV unparticle couplings.
The tensor unparticle contribution will not be considered as it is
suppressed by the inverse of $\Lambda_\U^2$, which stems from the fact that
the spin-2 unparticle operator is of higher dimension than that of the spin-0 and spin-1
unparticles. The study of $\mu-\tau$ conversion induced  by LFV scalar- and vector-mediated,
four fermion  couplings, $\bar\tau \mu
\bar q q$, via DIS has already been discussed in the context of effective Lagrangians
\cite{Sher:2003vi} and the
minimal supersymmetric standard model (MSSM) \cite{Kanemura:2004jt}. It was concluded that a 50 GeV
muon beam with intensity of $10^{20}$ muons on a nucleon target per year, as expected in a neutrino
factory
\cite{NF:2011aa}, would allow for about $10^6-10^7$  $\mu   N \to \tau  X$ events annually as long
as
the corresponding cross section is of the order of a few fb. This rate could give some room for
either detecting the signal or  placing stringent limits on $\mu-\tau$ couplings.

The rest of the work is organized as follows. Section II is devoted
to the calculation of the process $\mu  N \to \tau  X$ in the context
of unparticle physics. The numerical analysis and discussion is presented in
Sec. III. Finally, the conclusions and outlook are presented in Sec. IV.

\section{$\mu N\to \tau X$ cross section from unparticle interactions}

\begin{figure}[!ht]
\begin{center}
\includegraphics[width=3in]{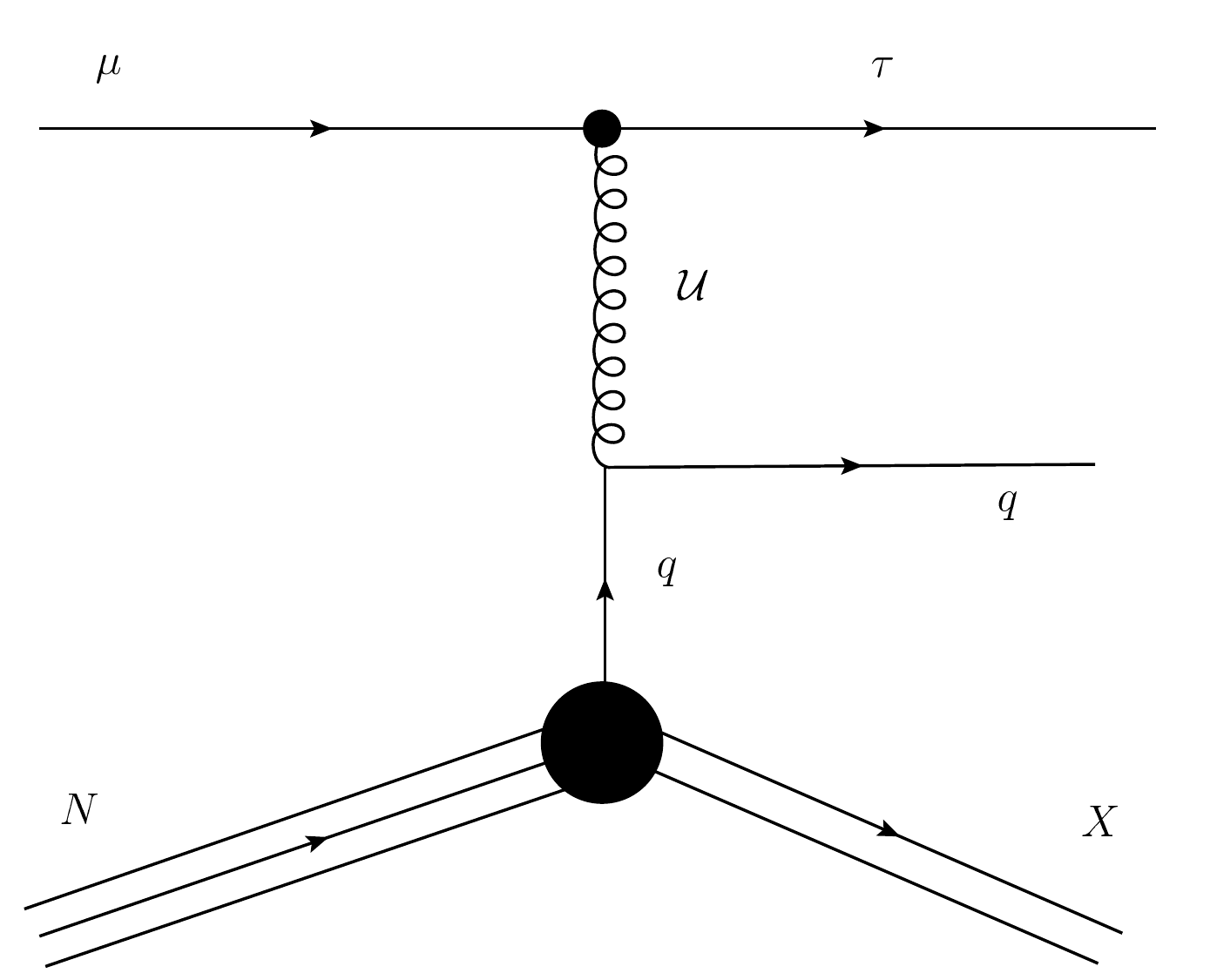}
\end{center}
\caption{Feynman diagram for $\mu N \to \tau X$ scattering due to
LFV unparticle couplings. We neglect flavor change in the quark sector.\label{vertex}}
\end{figure}

We will consider the DIS process $\mu N\to \tau X$  due to lepton flavor
violating  unparticle interactions,
which arises through the Feynman diagram of Figure \ref{vertex}. We are neglecting flavor
changing transitions in the quark sector. The most general
renormalizable  effective operators for the couplings of  spin-0 and spin-1 unparticles
to a fermion pair are
\cite{Georgi:2007ek}:
\begin{eqnarray}
{\cal L}_{{\cal U}^0} &=& \frac{\lambda_{{S}}^{ij}}{\Lambda_{{\cal
U}}^{d_{{\cal U}-1}}} {\bar f_i} f_j {\cal O}_{{\cal U}^0}
+\frac{\lambda_{{P}}^{ij}}{\Lambda_{{\cal U}}^{d_{{\cal U}-1}}} {\bar f_i}
\gamma^5 f_j {\cal O}_{{\cal U}^0}
,\label{eqn:lagsca}
\\
{\cal L}_{{\cal U}^1} &=& \frac{\lambda_{{V}}^{ij}}{\Lambda_{{\cal U}}^{d_{{\cal
U}-1}}} {\bar f_i} \gamma_{\mu} f_j {\cal O}^{\mu}_{{\cal U}^1}
+\frac{\lambda_{{A}}^{ij}}{\Lambda_{{\cal U}}^{d_{{\cal U}-1}}} {\bar f_i}
\gamma_{\mu} \gamma^5 f_j
{\cal O}^{\mu}_{{\cal U}^1}, \label{eqn:lagvec}
\end{eqnarray}
where $i$ and $j$ are flavor indexes. For the flavor diagonal couplings we will adopt the notation
$\lambda_J^{ii}\equiv\lambda_J^i$, with $J=S$, $P$, $A$, and $V$.

We will neglect all the fermion masses and calculate the unpolarized double differential cross
section for the
constituent parton subprocesses in terms of the usual $x$
and $y$ variables, where $x=Q^2/(2m_N\nu)$ is the fractional longitudinal momentum
carried by the struck parton and $y=\nu/E_\mu$ is the fractional energy transfer. Here
$Q^2$ is the squared momentum transfer and $\nu=E_\mu-E_\tau$, with $E_\mu$ and $E_\tau$
the muon and tau energies in the nucleon rest frame. In terms of the Mandelstam
variables of the parton subprocess we have $Q^2=-\hat t\simeq y \hat s=x y s$, with $s=2m_N E_\mu$
the square of the center-of-mass energy of the muon-nucleon collision. For completeness, we first
write the most
general expressions for the contributions of the spin-0 and spin-1 unparticles to the unpolarized
double differential cross section for the subprocesses $\mu q\to \tau q$ and $\mu \bar{q}\to \tau
\bar{q}$. For the spin-0 unparticle contribution we obtain

\begin{eqnarray}
\frac{d^2 \sigma_{\U^0}}{dx dy}(\mu q\to \tau q)=\frac{\hat s|\Delta_F(-Q^2)|^2}{16\pi
\Lambda_{\U}^{4(\du-1)}}
\left(|L_{\U^0}^{\mu\tau}|^2+|R_{\U^0}^{\mu\tau}|^2\right)\left(|L_{\U^0}^q
|^2+|R_{\U^0}^q|^2\right)y^2.
\end{eqnarray}
 which is also valid for the $\mu \bar q\to \tau \bar q$ subprocess. We have adopted the following
shorthand notation for the unparticle couplings:
$L_{\U^0}^{ij}=\lambda^{ij}_S-\lambda^{ij}_P$ and
$R_{\U^0}^{ij}=\lambda^{ij}_S+\lambda^{ij}_P$, whereas $\Delta_F(p^2)$ is given in Eq.
(\ref{scapropagator}).

On the other hand,
the corresponding contributions of a spin-1
unparticle are given by
\begin{eqnarray}
\frac{d^2 \sigma_{\U^1}}{dx dy}(\mu q\to \tau
q)&=&\frac{\hat s|\Delta_F(-Q^2)|^2}{64\pi\Lambda_{\U}^{4(\du-1)}}
\left(|L_{\U^1}^q|^2\left(|L_{\U^1}^{\mu\tau}|^2+|R_{\U^1}^{\mu\tau} |^2(1-y)^2\right)\right.
\nonumber\\&+&\left.|R_ { \U^1
}^q|^2\left(|R_{\U^1}^{\mu\tau}|^2+|L_{\U^1}^{\mu\tau}|^2(1-y)^2\right)\right),
\end{eqnarray}
and
\begin{eqnarray}
\frac{d^2 \sigma_{\U^1}}{dx dy}(\mu \bar q\to \tau \bar q)&=&\frac{\hat s|\Delta_F(-Q^2)|^2}{16\pi
\Lambda_{\U}^{4(\du-1)}}
\left(|L_{\U^1}^q|^2\left(|R_{\U^1}^{\mu\tau}|^2+|L_{\U^1}^{\mu\tau} |^2(1-y)^2\right)\right.
\nonumber\\&+&\left|R_
{ \U^1 }^q|^2\left(|L_{\U^1}^{\mu\tau}|^2+|R_{\U^1}^{\mu\tau}|^2(1-y)^2\right)\right).
\end{eqnarray}
with
$L_{\U^1}^{ij}=\lambda^{ij}_V-\lambda^{ij}_A$ and
$R_{\U^1}^{ij}=\lambda^{ij}_V+\lambda^{ij}_A$.

We can now fold the above expressions with the nucleon parton distribution
functions to obtain the cross section for the process $\mu N\to \tau X$. For instance, if
we consider an isoscalar nucleon $N=(n+p)/2$, with mass $m_N=(m_n+m_p)/2$, we obtain for the
contribution of a spin-0 unparticle:
\begin{eqnarray}
\frac{d^2 \sigma_{\U^0}}{dx dy}(\mu N\to \tau X)=\frac{m_N E_\mu}{32\pi \Lambda_{\U}^{4(\du-1)}}
|\Delta_F(-Q^2)|^2 x\left(q_{\U^0}(x,Q^2)+\bar{q}_{\U^0}(x,Q^2)\right) y^2,
\end{eqnarray}
and for the contribution of a spin-1 unparticle:
\begin{eqnarray}
\frac{d^2 \sigma_{\U^1}}{dx dy}(\mu N\to \tau X)&=&\frac{m_N E_\mu}{8\pi
\Lambda_{\U}^{4(\du-1)}}|\Delta_F(-Q^2)|^2 x
\left(q_{\U^1}(x,Q^2)+\bar{q}_{\U^1}(x,Q^2)(1-y)^2\right).
\end{eqnarray}
where
\begin{eqnarray}
q_{\U^0}(x,Q^2)=\bar{q}_{\U^0}(x,Q^2)=\left(|L_{\U^0}^{\mu\tau}|^2+|R_{\U^0}^{\mu\tau}
|^2\right)\left(|L_ { \U^0}^q
|^2+|R_{\U^0}^q|^2\right)\left(\frac{u_v+d_v}{2}+S
\right),
\end{eqnarray}
\begin{eqnarray}
q_{\U^1}(x,Q^2)&=&\left(|L_{\U^1}^{\mu\tau}|^2|L^q_{\U^1}|^2+
|R_{\U^1}^{\mu\tau}|^2|R^q_{\U^1}|^2\right)\left(u_v+d_v
\right)+
\left(|L_{\U^1}^{\mu\tau}|^2+|R_{\U^1}^{\mu\tau}|^2\right)\left(|L^q_{\U^1}
|^2+|R^q_{\U^1}|^2\right)S,
\end{eqnarray}
and
\begin{eqnarray}
\bar{q}_{\U^1}(x,Q^2)&=&\left(|L_{\U^1}^{\mu\tau}|^2|R^q_{\U^1}|^2+
|R_{\U^1}^{\mu\tau}|^2|L^q_{\U^1}|^2\right)\left(u_v+d_v
\right)+
\left(|L_{\U^1}^{\mu\tau}|^2+|R_{\U^1}^{\mu\tau}|^2\right)\left(|L^q_{\U^1}
|^2+|R^q_{\U^1}|^2\right)\bar{S}.
\end{eqnarray}
We have considered that the unparticle couplings to quark-antiquark pairs are flavor blind:
${L}_{\U^{0,1}}^u={L}_{\U^{0,1}}^d={L}_{\U^{0,1}}^q$,
${R}_{\U^{0,1}}^u={R}_{\U^{0,1}}^d={R}_{\U^{0,1}}^q$. Also,  $u_v$ and $d_v$ stand for the valence
quark  distribution functions, whereas
$S=\bar{S}=u_s+d_s+c_s+b_s+t_s$ stands for the
sea quark distribution function.  We omitted the explicit dependence on $x$
and $Q^2$.

Unfortunately there is dependence on several free parameters.
So, without losing generality we will assume that the pseudoscalar and vector-axial unparticle
couplings are negligible as
compared to the scalar and vector couplings, i.e.
$L_{\U^{0}}^{ij}\simeq R_{\U^{0}}^{ij}\simeq \lambda_S^{ij}$ and
$L_{\U^{1}}^{ij}\simeq R_{\U^{1}}^{ij}\simeq \lambda_V^{ij}$. In fact,  as
discussed in Ref. \cite{Moyotl:2011yv}, the contributions of the LFV couplings
$\lambda_{P,A}^{\mu\tau}$ to
the muon MDM, $a_\mu$, are negative and thus they are strongly disfavored by the current
experimental data \cite{Beringer:1900zz}, which require a positive contribution to $a_\mu$
to bring the theoretical prediction closer to the experimental value.  With these assumptions, we
obtain the
following
expressions for the double differential cross sections

\begin{equation}
\frac{d^2 \sigma_{\U^0}}{dx dy }(\mu N\to \tau
X)=\frac{m_N E_\mu}{8\pi\Lambda_{\U}^{4(\du-1)}} |\Delta_F(-Q^2)|^2 x\,q(x,Q^2)
 |\lambda_{S}^{q}|^2 |\lambda _{S}^{\mu\tau}|^2  y^2,
\label{difcross-0}
\end{equation}

\begin{eqnarray}
\frac{d^2 \sigma_{\U^1}}{dx dy }(\mu N\to \tau
X)=\frac{m_N E_\mu}{4\pi
\Lambda_{\U}^{4(\du-1)}}|\Delta_F(-Q^2)|^2x \,q(x,Q^2) |\lambda
_{V}^{q}|^2|\lambda
_{V}^{\mu\tau}|^2
\left(1 +(1-y)^2
\right).
\label{difcross-1}
\end{eqnarray}
with $q(x,Q^2)=u_v+d_v+2S$.
Therefore the $\mu N \to \tau
X$ cross section due to spin-0 and spin-1 unparticle exchange is

\begin{equation}
\sigma(\mu N \to \tau X)= \sum_{i=0,1}\int_0^1 \int_0^1
\frac{d^2\sigma_{\U^i}}{dx dy }(\mu N\to \tau X)dx dy.\label{sigma}
\end{equation}

It is also useful to express the double differential cross sections (\ref{difcross-0}) and
(\ref{difcross-1}) as functions of the angle of the tau
lepton with respect to the beam direction and the tau energy. In
terms of these variables, we have $x=Q^2/(2m_N(E_\mu-E_\tau))$ and
$y=(E_\mu-E_\tau)/E_\mu$, with $Q^2=2 E_\mu E_\tau(1-\cos\theta)$. The
transformation from the $(x,y)$ variables to
 $(E_\tau,\theta)$ can be written as

\begin{equation}
\frac{d^2\sigma_\U}{dE_\tau d\theta}(\mu N\to\tau X)= J(E_\tau,\theta)
\frac{d^2\sigma_\U}{dx dy}(\mu N\to\tau X).
\end{equation}
where  the Jacobian of the
transformation from the $(x,y)$ variables to
the $(E_\tau,\theta)$ variables is $J(E_\tau,\theta)= E_\tau \sin\theta/(m_N (E_\mu-E_\tau))$.

Before the numerical evaluation of Eq. (\ref{sigma}), we need to
discuss the current bounds on the fermion unparticle couplings and the unparticle scale and
dimension. We will assume that the
unparticle couplings to quark pairs are flavor blind, with $\lambda^q_{S,V}\simeq O(1)$, whereas for
the
LFV couplings $\lambda_{S,V}^{\mu\tau}$  we will consider the most stringent constraints obtained
from the experimental limits on LFV tau decays and the muon MDM for values of $\Lambda_\U$ and $\du$
consistent with the search for
monojets plus missing transverse energy at the LHC by the CMS collaboration
\cite{Chatrchyan:2011nd}.

\section{Numerical results and discussion}

\subsection{Constraints on unparticle couplings}

Shortly after the advent of Georgi's unparticle conjecture, bounds on the scale
$\Lambda_\U$ and the dimension $d_\U$ were obtained by using the LEP data of monophoton
production plus missing transverse energy \cite{Cheung:2007ap}. More recently, the data of the
search
for monojet production plus missing transverse energy at the LHC were used by the CMS collaboration
to constrain the parameters associated with a spin-0  unparticle
\cite{Chatrchyan:2011nd}. It was concluded that the region $\du\leq 1.4$ is strongly
disfavored as $\Lambda_\U\ge 10$ TeV is required to be
consistent with the LHC data. On the other hand, for $\du$ close to 2,  $\Lambda_\U\simeq 1$ TeV is
still  allowed. We will thus consider
three illustrative sets of $(d_\U,\Lambda_\U)$ values consistent with the CMS
bounds, namely, ($1.4$, $10$ TeV), ($1.6$, $5$ TeV), and
($1.9$, $1$ TeV). It is worth noting that the CMS bounds were obtained
assuming unparticle couplings of the order of unity. So, these bounds would be weaker if couplings
of smaller size were considered.

We now turn to discuss the current constraints on the LFV unparticle couplings. As stated above, for
simplicity we will neglect FCNC unparticle couplings in the quark sector and consider that the
diagonal couplings $\lambda_{S,V}^q$ are flavor blind and of the order of $O(1)$. This is in
accordance, for instance,  with the conclusions reached in  Ref. \cite{Chen:2008ie}, where a study
of the effects of a vector unparticle on the  $B\to \pi\pi$ and $B\to \pi K$ decays, combined with
the constraints on $B_{d,s}-\bar{B}_{d,s}$ mixing, was presented. It was found that, for $d_{\cal
U}=1.5$, a minimum $\chi^2$ analysis yields that the contribution of a vector unparticle can be in
agreement with all the measurements of the $B\to \pi\pi$ and $B\to \pi K$ decays as long as the
$\lambda_V^{u}$ and $\lambda_V^{d}$ couplings are of the order of $O(1)$, with both the
$\lambda_V^{sb}$ and $\lambda_V^{db}$ couplings being of the order of $10^{-4}$.

As far as the LFV unparticle  couplings are concerned, they can be constrained from the experimental
data of
the muon MDM and the LFV decays $ l_i\to  l_j  l_k   l_k$ and $ l_i\to  l_j \gamma$, with $l_{i,j}$
a charged lepton. In addition,  the experimental limits on  the semileptonic decays $\tau\to l_i
M_i$ and $\tau\to l_i M_i M_j$ \cite{Beringer:1900zz}, with $M_{i,j}$ a generic light meson,   can
be useful to put stringent constraints on the tau LFV couplings
\cite{Black:2002wh,Brignole:2004ah,Sher:2002ew,Fukuyama:2005bh}. We will start by discussing the
constraints
obtained from the leptonic tau decay channels. By using the experimental data on the muon MDM and
the $\tau\to3\mu$ decay,  the allowed region in the
$\lambda_{S}^{\mu\tau}$ vs $\lambda_S^{\mu\mu}$ plane  was obtained in \cite{Hektor:2008xu} for
several values of $\du$ and
$\Lambda_\U$. A similar procedure was used in \cite{Moyotl:2011yv}  to obtain the allowed area in
the  $\lambda_V^{\mu\tau}$ vs  $\lambda_V^{\mu\mu}$ plane. It was also found that the loop-induced
decay $l_j\to l_i\gamma$ gives a weaker constraint on such unparticle couplings.
For the three sets of
$(d_\U,\Lambda_\U)$ values chosen above, we show in  Table \ref{table:bounds} the maximal
allowed values of the $\lambda_{S,V}^{\mu\tau}$ couplings along with the corresponding values of the
$\lambda_{S,V}^{\mu\mu}$ coupling. In general,  $\lambda_V^{\mu\tau}$ is
more
constrained than
$\lambda_S^{\mu\tau}$, but both couplings become tightly constrained when $d_\U$ gets closer to
1.4. For more details of these analyses, we refer the interested reader to
the original references \cite{Hektor:2008xu,Moyotl:2011yv}. We also would like to comment on the
bounds  obtained from the tau semileptonic decay channels. Contrary to the bounds obtained from the
leptonic tau decays, in which only the  couplings to lepton pairs are involved, both the couplings
to lepton pairs and quark pairs enter into the semileptonic tau decay amplitudes.
Under our assumptions that the nondiagonal unparticle couplings to quarks are much smaller than the
diagonal ones, which we assume to be of the order of unity, the two-body decays $\tau \to \mu \pi^0$
and $\tau \to \mu \eta$  could be useful to constrain the LFV pseudoscalar  $\lambda_P^{\mu\tau}$
and
axial vector $\lambda_A^{\mu\tau}$ couplings, whereas the LFV vector  $\lambda_V^{\mu\tau}$ coupling
could be constrained via the $\tau \to \mu \rho$ and $\tau\to\mu \phi$ decays. On the other hand,
the experimental limits on the three-body decays $\tau \to \mu \pi^0\pi^0$, $\tau\to\mu \eta \eta$,
$\tau\to \mu\pi^- \pi^+$, etc.  can translate into bounds on the scalar $\lambda_S^{\mu\tau}$
coupling. Along these lines, the authors of Ref. \cite{Li:2009yr} studied the constraints on LFV
vector unparticle couplings  from the decays $\tau\to \mu V^0$ ($V^0=\rho, \omega, \phi$). However,
it was concluded that the resulting constraints turn out to be weaker than the constraints obtained
from the muon MDM and the  leptonic tau decays for    $\Lambda_{\cal U}=1$ TeV, $1.6\le d_{\cal
U}\le 2$, and values of the unparticle couplings to quark pairs in the interval $0.1-1$
\cite{Li:2009yr}. We will
thus  consider the restrictions of Table \ref{table:bounds} in the following analysis.

In summary, to illustrate the behavior of the $\mu N\to \tau X$ cross section, we will consider the
maximal allowed
values of the $\lambda_{S,V}^{\mu\tau}$ couplings consistent with the
current bounds on the muon MDM and the LFV decay $\tau\to
3\mu$, for three sets of $(d_\U,\Lambda_\U)$ values   consistent with the bounds obtained by the CMS
collaboration.

\begin{table}[!ht]
\begin{tabular}{llllll}
\hline
$\Lambda_{\cal U}$ (TeV) & $d_{\cal U}$&$\lambda_S^{\mu\tau}$
&$\lambda_S^{\mu\mu}$&$\lambda_V^{\mu\tau}$ &$\lambda_V^{\mu\mu}$
\\
\hline
$10$ & $1.4$&$4\times 10^{-3}$&$5 \times 10^{-2}$&$5\times 10^{-4}$&$8 \times 10^{-2}$ \\
$5$ & $1.6$&$1\times 10^{-2}$&$0.4$&$2\times 10^{-3}$&$0.6$ \\
$1$ & $1.9$&$3\times 10^{-2}$ &$1.2$&$5\times 10^{-3}$&$4.5$\\
\hline
\end{tabular}
\caption{$95\%$ C. L. upper limits on the unparticle couplings $\lambda_V^{\mu\tau}$ and
$\lambda_S^{\mu\tau}$, from the current bounds on the MDM and the
LFV decay $\tau\to 3\mu$, for three sets of $(\du,\Lambda_\U)$ values consistent with the bounds
obtained
by the CMS collaboration
\cite{Chatrchyan:2011nd}. The corresponding values of the $\lambda_{S,V}^{\mu\mu}$ couplings are
also shown.}\label{table:bounds}
\end{table}
\subsection{Unparticle contribution to the $\mu P\to \tau X$ cross section}

\begin{figure}[!ht]
\begin{center}
\includegraphics[width=6in]{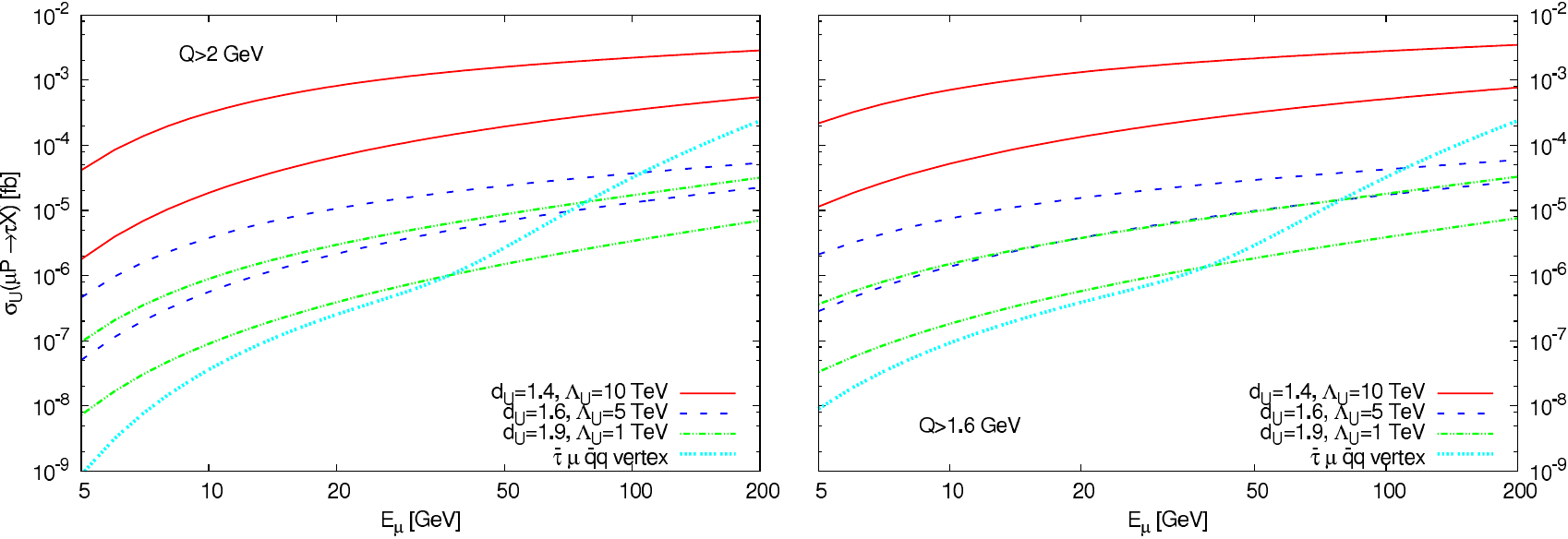}
\end{center}
\caption{ $\mu P \to \tau X$ cross section due to
LFV unparticle couplings for
three sets of $(\du, \Lambda_\U)$ values. For the LFV coupling $\lambda_{S,V}^{\mu\tau}$ we used the
values
shown in  Table \ref{table:bounds}, whereas $\lambda_{V,S}^q\simeq O(1)$ was assumed for
all the unparticle-quark couplings.  A cut of $Q>2$ ($1.6$) GeV was used in the left (right) plot.
For each line style, the upper lines correspond
to the contribution from a spin-0 unparticle, whereas the lower lines represent the contribution
from a
spin-1 unparticle. For comparison purposes, we also show the
contribution from a
dimension-six effective four-fermion LFV vertex $\bar{\tau}\mu \bar{q}q$, considering the coupling
values used in \cite{Kanemura:2004jt} for the contribution of
Higgs exchange in the MSSM. \label{cross.section}}
\end{figure}

We now turn to the numerical analysis. We will consider
that the target is a proton and use the CTEQ6m parton distribution functions
\cite{Pumplin:2002vw}. We show  the spin-0 and spin-1 unparticle contributions to
the $\mu P\to \tau X$ cross
section in Fig. \ref{cross.section} as a function of the muon energy and for three sets of
$(d_\U$,$\Lambda_\U$) values. For each such set,
we used the
corresponding maximal allowed values of $\lambda_{S}^{\mu\tau}$ and
$\lambda_{V}^{\mu\tau}$  shown in  Table \ref{table:bounds}. In Fig. \ref{cross.section} we present
two plots in
which we
consider a different cut in the momentum transfer: in the left plot we take $Q>2$ GeV, whereas in
the right plot we use $Q>1.6$ GeV. Since
$\lambda_{S,V}^q\simeq O(1)$ was assumed
for all the quarks, it is worth noting that if this coupling was decreased by  one order of
magnitude, the cross section would decrease by two orders of magnitude. For comparison
purposes, we have also included in these plots the contribution to the  $\mu P\to \tau X$ cross
section from the
dimension-six effective scalar-mediated four-fermion LFV vertex $\bar{\tau}\mu \bar{q}q$, which was
already studied by the authors of Ref. \cite{Kanemura:2004jt}, who focused on the contribution of
Higgs exchange  in  the context of
the minimal supersymmetric standard model (MSSM).  We note that our results
agree with those presented in \cite{Kanemura:2004jt}, which serves as a cross-check
for our calculation.

It is interesting that  the $\mu P\to \tau X$ cross section is larger when  $\du=1.4$
and smaller when  $\du$ approaches 2, which contrasts with the size of the values assumed for the
LFV unparticle coupling $\lambda_{S,V}^{\mu\tau}$ and the unparticle scale $\Lambda_\U$: according
to the aforementioned bounds, when $\du=1.4$,
$\lambda_{S,V}^{\mu\tau}$ is smaller and  $\Lambda_\U$ is larger, but when $\du=1.9$,
$\lambda_{S,V}^{\mu\tau}$ is larger and $\Lambda_\U$ is smaller.
The results observed in Fig. \ref{cross.section} stem from  the behavior of the unparticle
propagator [Eq. (\ref{scapropagator})]: the unparticle
contribution behaves as that of a massless  particle as $\du\to 1$, but it approaches
the contribution of a four-fermion contact vertex when $\du\to 2$. We also note that the spin-0
unparticle
contributions  are larger by more than one order of magnitude than
the spin-1 unparticle contributions, which is a result of  the values used  for
the $\lambda_{V}^{\mu\tau}$ and $\lambda_{S}^{\mu\tau}$ couplings. This situation is also
observed in the case of the scalar-mediated and the vector-mediated contributions
studied in \cite{Sher:2003vi,Kanemura:2004jt}. Finally, although both the
spin-0 and the spin-1 unparticle contributions
increase steadily with $E_\mu$, such increase is not as dramatic as it does occur in the case of the
MSSM contribution.

Figure \ref{cross.section} also shows the sensitivity of the $\mu P\to \tau X$ cross section to the
cut in the momentum transfer. This is also a reflect of the infrared behavior of the unparticle
propagator. Even if the
cut $Q>2$ GeV is imposed,  for a wide range of $E_\mu$ values, the unparticle
contributions  can be larger than the
MSSM contribution, though the latter  increases suddenly
around $E_\mu=50$ GeV  and continues to increase steadily. As
explained in \cite{Kanemura:2004jt}, such a  dramatic increase is due to the contribution of the sea
$b$ quark, which is enhanced by a factor of
$(m_b/m_s)^2$ with respect to that of the sea $s$ quark. On the other hand, the
unparticle contributions
are considerably smaller than those obtained in \cite{Sher:2003vi} for the calculation of the
quasielastic process $\mu N\to \tau N$ using the four-fermion LFV scalar coupling. However, in
that analysis, a value of $4\pi/\Lambda^2$, with $\Lambda=1$ TeV, was used for the associated
coupling constant, whereas we
are
considering strong bounds on the LFV unparticle coupling constants.

\subsection{Angular and energy distributions of the emitted tau lepton}

\begin{figure}[!ht]
\begin{center}
\includegraphics[width=6in]{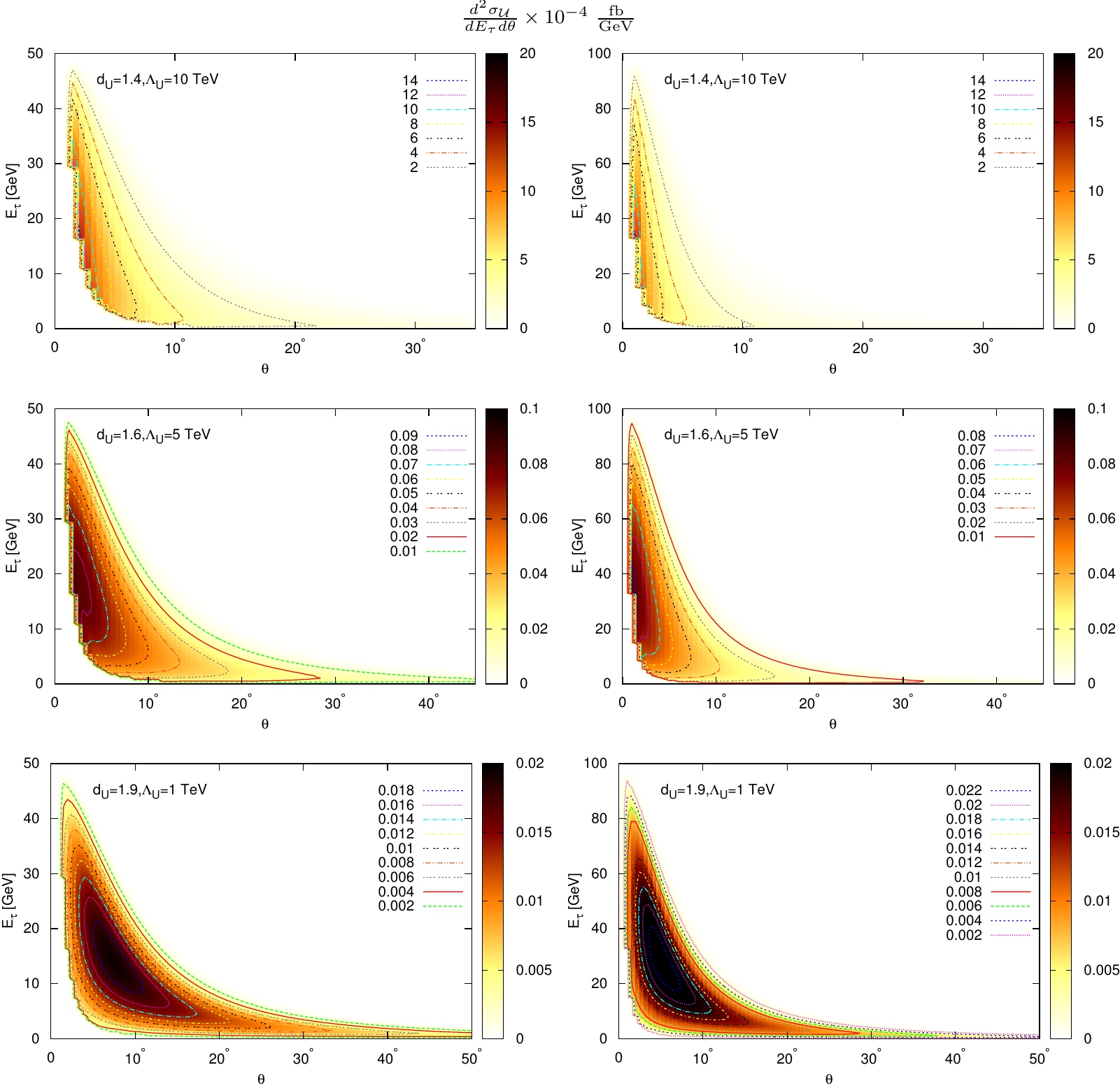}
\end{center}
\caption{Contour plots for the spin-0 unparticle-mediated double differential cross section
$\frac{d^2\sigma}{dE_\tau
d\theta}(\mu P\to\tau X)$  for $E_\mu=50$ GeV (left plot) and
$E_\mu=100$
GeV (right plot) for the parameters of Table
\ref{table:bounds}. The cut $Q>2$ GeV is imposed.\label{contourplot}}
\end{figure}

We now would like to discuss the behavior of the angular and energy distributions of the emitted
tau lepton for our DIS process, which could be a useful tool to disentangle
the unparticle contributions from other class of effects. For this purpose, in Fig.
\ref{contourplot} we show the contour lines of the spin-0 unparticle-mediated double
differential cross section $\frac{d^2\sigma_\U}{dE_\tau d\theta}(\mu P\to \tau X)$
for the same sets of  parameter values of Table \ref{table:bounds} and two values of the
muon energy. The analogous plots for the spin-1
unparticle-mediated contribution show a similar
behavior and we refrain from presenting them here. It is worth noting that the cut $Q>2$ GeV was
used in these plots, which explains
the white area closer to $\theta=0$. In fact, in this area the double differential cross section
could reach its higher values depending on the $\du$ value, as will be discussed below. Also, our
calculation
automatically excludes the kinematically forbidden region, which appears as the unshaded area in
the top right corner of each plot. Due to the infrared behavior of the unparticle
propagator [Eq. (\ref{scapropagator})], it is expected that  the tau lepton would be
emitted preferentially along
the forward direction of the beam when $\du$ is close to $1$. This is shown in Fig.
\ref{contourplot}, where the darker area
on the contour plots, which is associated with
the higher values of the double differential cross section, spreads around $\theta=0$, i.e. the
higher values of the double differential cross section are reached around $\theta=0$. We can see
that the darker area is relatively wider at low tau energies, but it shrinks considerably as the tau
energy increases. It
means that, in this situation, a low-energy tau would be  emitted at a larger angle than a
high-energy
tau.
On the other hand,
the situation is rather different when $\du$ approaches 2, when the infrared behavior of the
unparticle propagator is less pronounced. We thus observe that, when $\du=1.9$, the darker
area in
the contour plot shrinks considerably and also  shifts rightward and upward. In this
case the
preferred emission angle of the tau lepton is no longer located close to  the forward beam direction
but at a slightly larger angle whose value increases as the muon energy increases.
Again, a low-energy tau would tend to be emitted at a larger angle than an energetic tau.

\begin{figure}[!ht]
\begin{center}
\includegraphics[width=6in]{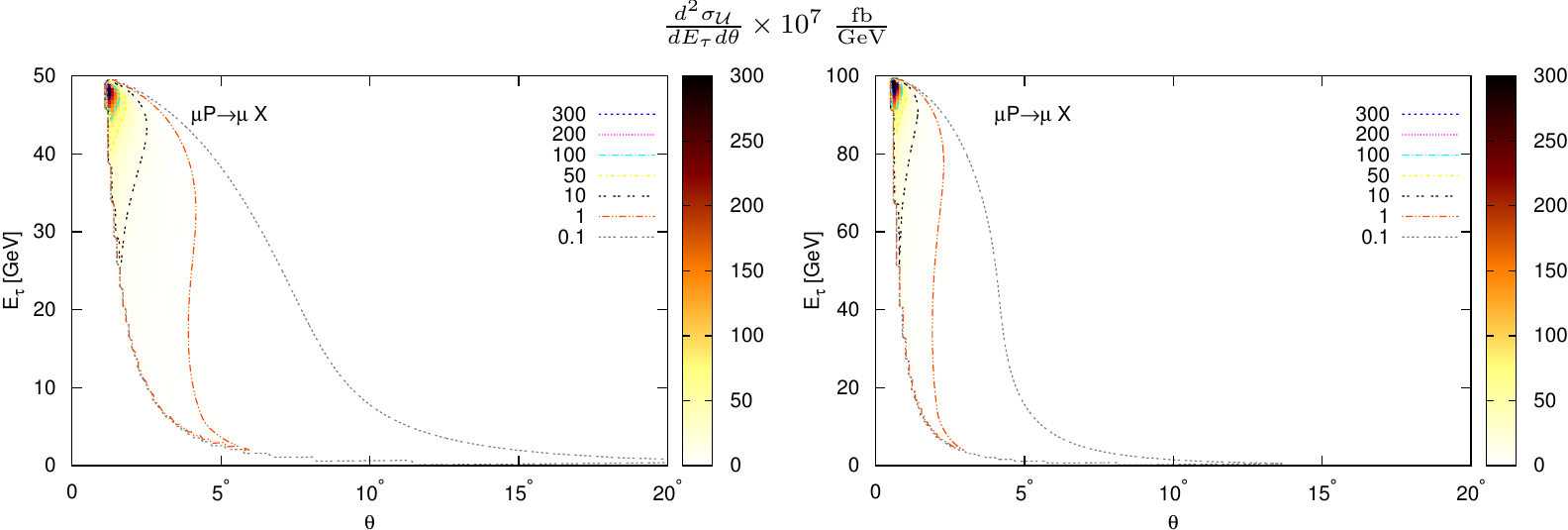}
\end{center}
\caption{Contour plots for the  double differential cross section
$\frac{d^2\sigma}{dE_\tau
d\theta}(\mu P\to\mu X)$   for $E_\mu=50$ GeV (left plot) and
$E_\mu=100$
GeV (right plot). The cut $Q>2$ GeV is imposed.\label{contourQplot}}
\end{figure}

\begin{figure}[!ht]
\begin{center}
\includegraphics[width=6in]{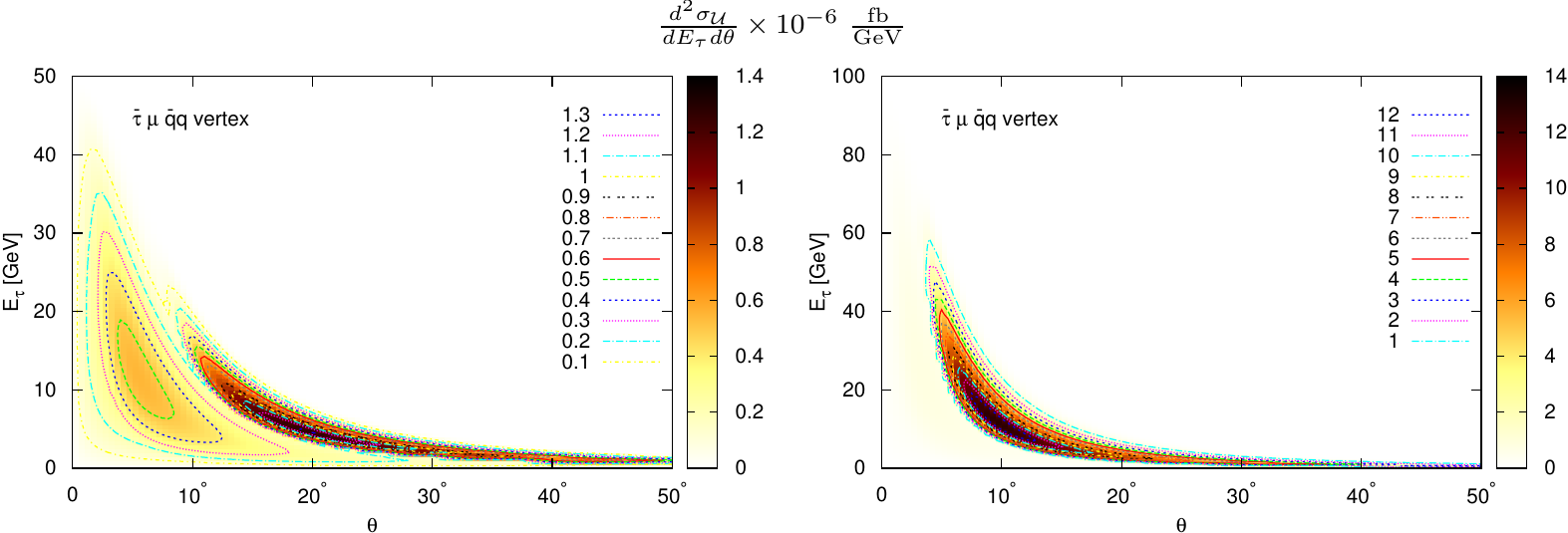}
\end{center}
\caption{Contour plots for the Higgs-mediated double differential cross section
$\frac{d^2\sigma}{dE_\tau
d\theta}(\mu P\to\tau X)$  for $E_\mu=50$ GeV (left plot) and
$E_\mu=100$
GeV (right plot). We used  the same coupling
values used in \cite{Kanemura:2004jt}. No cut is imposed. \label{contourKplot}}
\end{figure}

In conclusion, when $d_\U=1.4$ we expect that the $\mu P\to \tau X$double differential cross
section behaves similarly to that of photon-mediated $\mu P\to \mu X$ DIS, which in fact would be
the main source of background for our process. On the other hand, when $d_\U=1.9$ the behavior of
the $\mu P\to
\tau X$ double differential cross section would resemble that of the Higgs-mediated one.
It is interesting to contrast the behavior of all these kinds of contributions. We
thus show in Fig. \ref{contourQplot} the photon-mediated
$\mu P\to \mu X$ double differential cross section, whereas the
 the Higgs-mediated $\mu P\to \tau X$ double differential cross section is shown in Fig.
\ref{contourKplot}. In the former case (Fig. \ref{contourQplot}) we observe that the double
differential cross section is  strongly peaked at a low  angle and a high energy: in this
scenario, the signature of the process would be
an energetic muon  emitted close to the forward beam direction. Therefore,
the photon-mediated
$\mu P\to \mu X$ is highly sensitive to a cut in the transfer momentum $Q$. As far as the
Higgs-mediated double differential cross section is concerned, we observe in Fig.
\ref{contourKplot} that the tau lepton would be emitted preferentially with a low
energy (below one half the muon energy) and at an angle considerably larger than that of the
forward beam
direction: around $20^\circ$ for $E_\tau=50$ GeV and around $10^\circ$ for $E_\tau=100$ GeV. It is
also
interesting to note that the Higgs-mediated double differential cross section shows two peaks,
though the higher peak appears at a larger angle.

\begin{figure}[!ht]
\begin{center}
\includegraphics[width=6in]{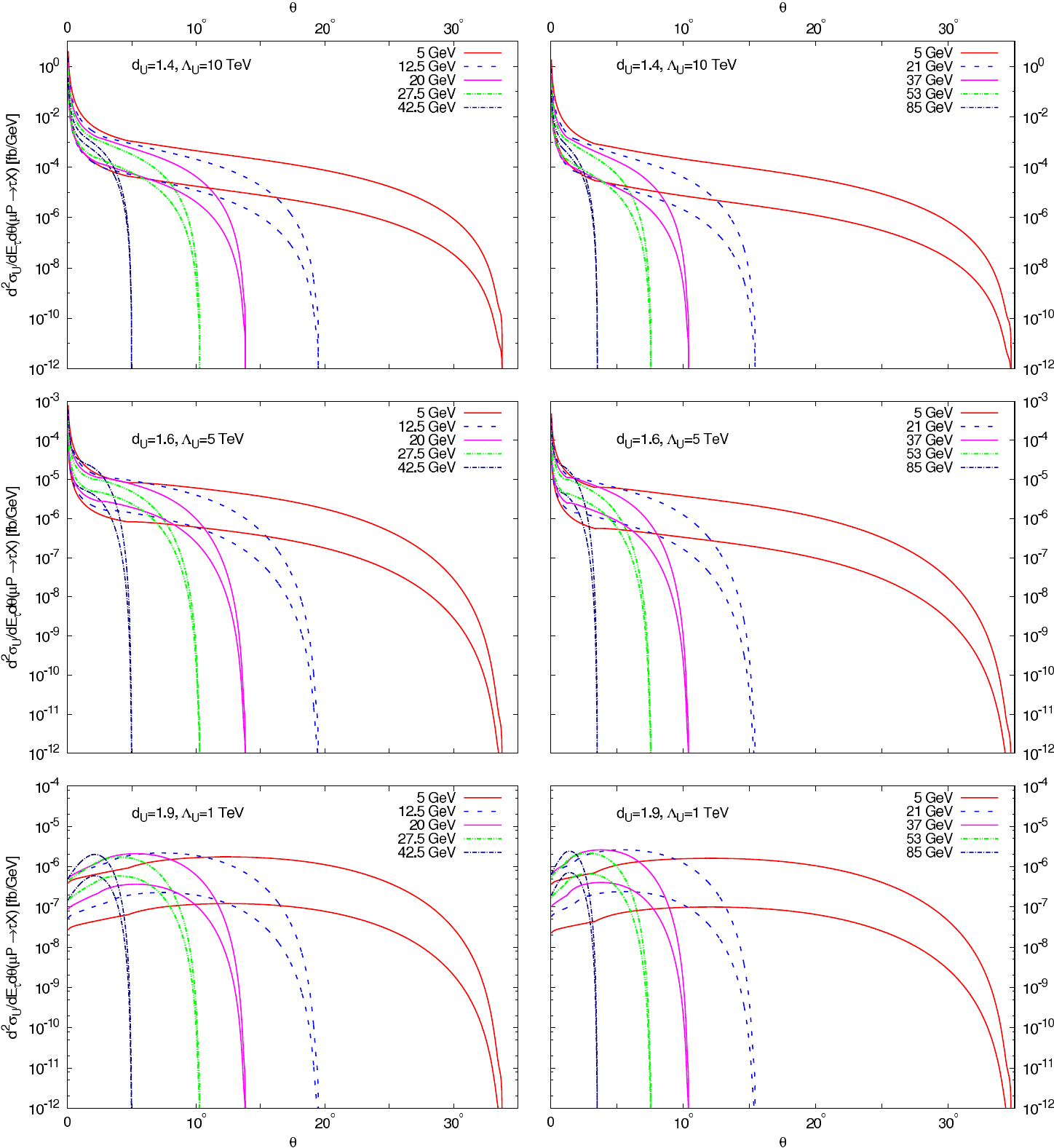}
\end{center}
\caption{Unparticle contribution to the double differential cross section $\frac{d^2\sigma}{dE_\tau
d\theta}(\mu P\to\tau X)$ as a function of the tau emission
angle $\theta$ for
several values of the tau energy $E_\tau$ and two values of $E_\mu$: 50 GeV (left plots) and 100
GeV (right plots). We
considered the three sets of parameter values of Table
\ref{table:bounds}.
For each line
style, the upper lines correspond to the spin-0 unparticle
contribution whereas the lower lines represent the spin-1 unparticle contribution.
\label{difcrossang}}
\end{figure}

More details of the behavior of the angular distribution of the tau lepton can be
extracted from Fig. \ref{difcrossang}, where,
for the same  parameter values used previously, we have plotted the
$\frac{d^2\sigma_\U}{dE_\tau d\theta}(\mu P\to \tau X)$ double differential cross section as a
function
of $\theta$, for several fixed values of $E_\tau$. We show plots for two values of $E_\mu$, namely,
50
GeV and 100 GeV. We can distinguish two cases: when $\du$ is close to 1 and when $\du$ is close to
2. In the first case, when $\du=1.4$, we can
observe that the tau lepton is emitted preferentially at angles
smaller than $35^\circ$, but the double differential cross section is considerably
larger around $\theta=0$, which means that the preferred emission angle of the tau lepton is
$\theta=0$. As its energy
increases, the tau lepton would tend to be emitted closer to the forward
direction of the beam. For instance, a tau lepton with about 90\%  the beam energy would be
emitted preferentially at $\theta\leq 5^\circ$, whereas a tau lepton with an energy about 10\%  the
beam energy would be emitted mainly at
$\theta\leq 35^\circ$.  The situation changes drastically when $\du$ is close to 2, namely
$\du=1.9$, in which case the tau lepton is emitted preferentially at large angles,
although the preferred angle gets closer to the
forward direction of the beam if the tau lepton energy is high.  In this
scenario, when $E_\tau$ is about 10\%  the beam energy, the preferred emission angle of the tau
lepton
is around 20$^\circ$, whereas it is around 5$^\circ$ when $E_\tau$ is about 90\% the beam energy.

\begin{figure}[!ht]
\begin{center}
\includegraphics[width=6in]{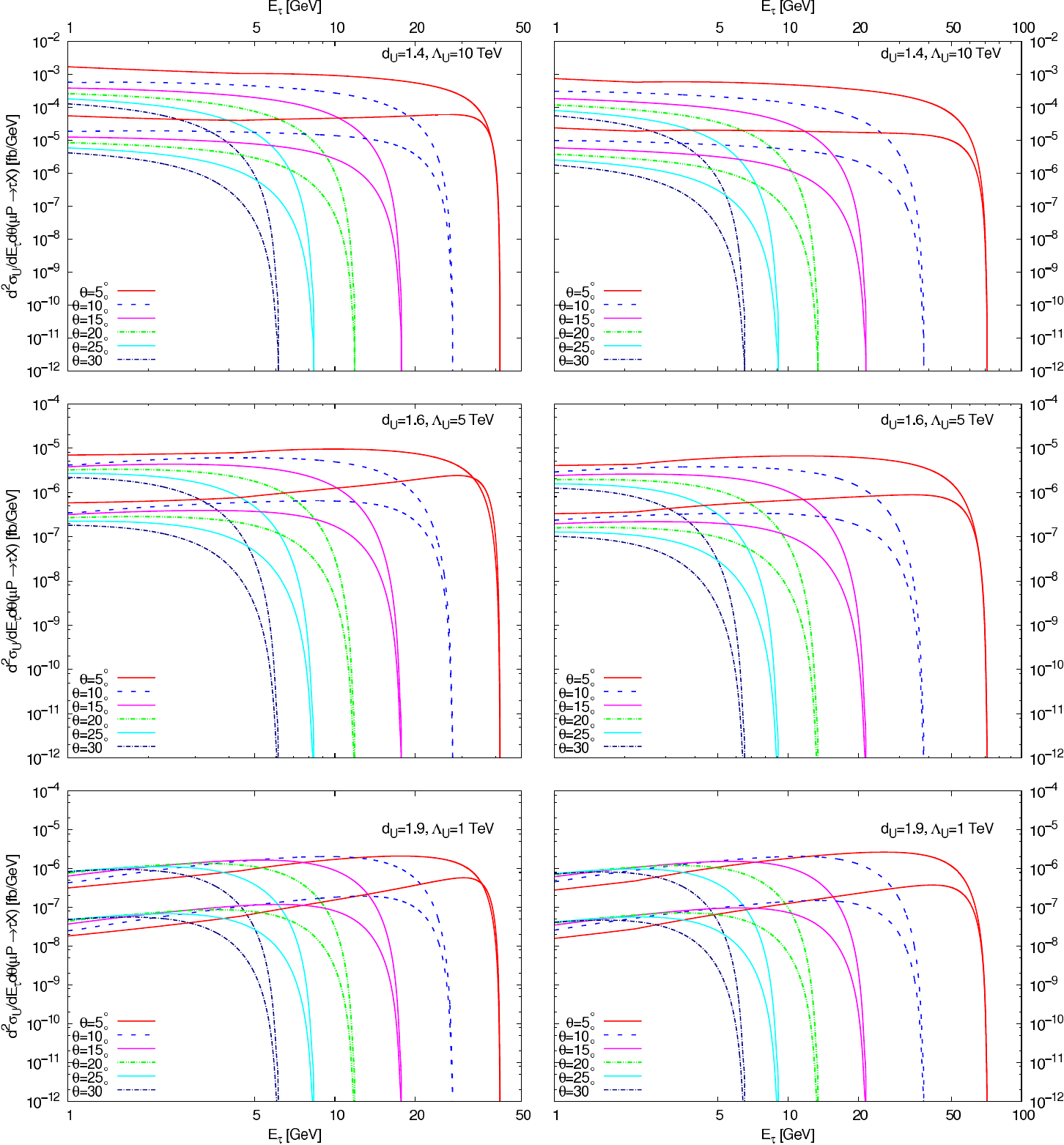}
\end{center}
\caption{Unparticle contribution to the double differential cross section $\frac{d^2\sigma}{dE_\tau
d\theta}(\mu P\to\tau X)$ as a function of the tau
energy $E_\tau$ for several values of the emission angle $\theta$ and two values
of $E_\mu$: 50 GeV (left plots) and 100
GeV (right plots). We considered
the three sets or parameter values of Table
\ref{table:bounds}. For each line style, the upper lines correspond to the spin-0 unparticle
contribution whereas the lower lines represent the spin-1 unparticle contribution.
\label{difcrossen}}
\end{figure}

The effects discussed above are also evident when we analyze the energy distribution of the tau
lepton. This is
shown in Fig. \ref{difcrossen}, where this
time we have plotted the double differential cross section as a function of $E_\tau$ for several
values of the emission angle $\theta$ and two values of $E_\mu$. We have used the
same set of parameter values used in the previous Figures. We observe that
the curves corresponding to increasing values of $\theta$ are shifted downward and leftward.
The area under each curve shrinks considerably as $\theta$ increases, which means that the bulk of
the contribution to the cross section arises mainly in the region of small angles, though this
situation changes slightly as $\du\to 2$. Although the
behavior of the differential cross section could be expected to be similar to that induced by a
massless
intermediary particle, the nature of the unparticle propagator  makes this effect
very distinctive as it is tuned by value of the dimension $\du$. We note, however, that all the
analysis we have done so far shows that there is
little difference between the contributions to
the $\mu P\to\tau X$ process from a spin-0 unparticle and from  a spin-1 unparticle, so unparticle
contributions would be hard to disentangle using this kind of analysis.

\begin{figure}[!ht]
\begin{center}
\includegraphics[width=6in]{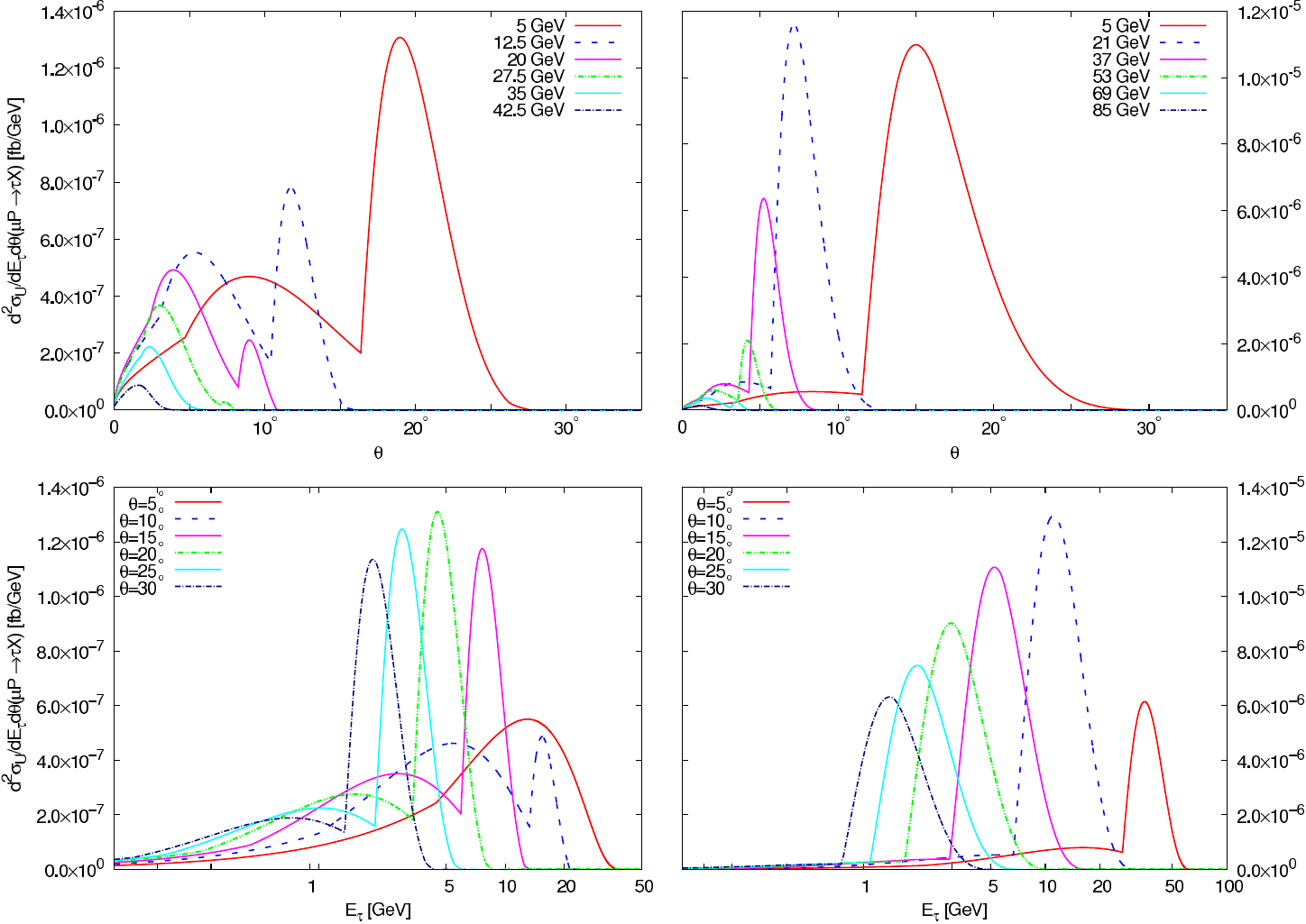}
\end{center}
\caption{Contribution to the double differential cross section $\frac{d^2\sigma}{dE_\tau
d\theta}(\mu P\to\tau X)$  from a
dimension-six effective four-fermion LFV vertex $\bar{\tau}\mu \bar{q}q$, considering the coupling
values used in \cite{Kanemura:2004jt}. The upper plots show the dependence on the tau scattering
angle $\theta$ for several values of the tau energy $E_\tau$,
whereas the lower plots show the
dependence on the tau energy for several values of the tau scattering angle. Two values of the muon
energy are used: $E_\mu=50$ GeV (left plots) and $E_\mu=100$
GeV (right plots).
\label{difcrosskane}}
\end{figure}

Finally, we would like to contrast the behavior of the angular and energy distributions of the
unparticle-mediated contribution to the
$\mu P\to\tau X$ process with that of the contribution from Higgs exchange
\cite{Kanemura:2004jt}. For comparison
purposes, we have made analogous plots to the ones shown in Figures \ref{difcrossang} and
\ref{difcrossen}  using the same parameter values as in Ref. \cite{Kanemura:2004jt}. The results are
presented in Fig. \ref{difcrosskane},
where we show the behavior of the scalar contribution to the $\mu P\to\tau X$ double differential
cross
section as a function of $\theta$ (upper plots) and $E_\tau$ (lower plots), for two values of
$E_\mu$. In this case a low-energy tau lepton is emitted preferentially at a
relatively large angle, whereas a high-energy tau lepton is emitted closer
to  the forward  direction of the beam. When the muon energy increases, the preferred emission angle
decreases. For instance, for $E_\mu=50$ GeV the largest peak in the double differential cross
section
is around
$\theta=20^\circ$, but for $E_\mu=100$ GeV the largest peak is around $\theta=10^\circ$.

\subsection{Background}

A beam with intensity of $10^{20}$ muons per year is expected at a neutrino factory, with the
muon energy in the range of a few dozens of GeVs \cite{NF:2011aa}. It has been estimated
\cite{Sher:2003vi} that, a cross section for the muon-nucleon collision of the order of 1 fb would
yield a probability of interactions per meter of about $6\times 10^{-14} \rho$ in a meter of target
as long as there is little ionization loss. Here $\rho$ is the density of the target expressed in
g/cm$^3$. Assuming $10^{20}$ muons per year on a $10^{2}$ g/cm$^2$ target mass  would yield about
 $10^{6}$ $\mu N \to \tau X$ events annually. The unparticle contributions to the $\mu P \to \tau X$
cross section are smaller than 1 fb but in a promising scenario we would have a cross
section of the
order of $10^{-4}-10^{-3}$ fb,  which would yield about $10^2-10^3$ $\mu P\to
\tau X$ events annually, though there would be some enhancement if
the target is an atom nucleus. The main issue for the detection of the signal of this reaction will
be the
identification of the tau lepton from its decay products. If the leptonic decay channel $\tau\to \mu
\bar{\nu}_\mu
\nu_\tau$ is considered, the most dangerous background is expected to arise from the lepton flavor
conserving reaction $\mu P\to \mu X$, which would proceed mainly via QED. For this reaction the muon
would also emerge dominantly along the beam forward direction, though its energy distribution would
be rather
different than that of the muon arising from the tau decay. A detailed discussion about reducing
this background can be found in \cite{Gninenko:2001id}. Other possibilities for detection of the
tau lepton has been examined in \cite{Kanemura:2004jt}, such as considering the hadronic tau decay
$\tau\to \pi \nu_\tau$. In this case, the main problem arises from the misidentification of
the pion with the muon arising from $\mu P\to \mu X$. A more detailed Monte Carlo
analysis would be required to make further conclusions.

\section{Conclusions}
We have studied $\mu-\tau$ conversion through the  $\mu P\to \tau X$ process  mediated
by spin-0 and spin-1 unparticles. For the model parameters, we used the most recent constraints on
the LFV unparticle couplings $\lambda_{S,V}^{\mu\tau}$ from the muon MDM and the tau decay
$\tau\to 3\mu$. These values are also consistent with the most recent bounds on
the unparticle scale $\Lambda_\U$ and the dimension $\du$ from the data of the search for monojets
plus
missing transverse energy at the LHC by the CMS collaboration. In a promising scenario, the
resulting cross section can be of the order of $10^{-3}-10^{-2}$ fb for $\du=1.4$ and $\Lambda_\U=10$
TeV. Due to the
infrared nature of the unparticle propagator, the angular distribution of the
emitted tau lepton is rather different than that observed in the case of other contributions: in the unparticle mediated process, the
tau lepton is emitted mainly along the forward beam direction. For a beam with intensity of
$10^{20}$
$50$ GeV muons per year on a target nucleon of $10^2$ gr/cm$^2$ mass, there would be about
$10^2-10^3$ $\mu P\to \tau X$ events annually, which would open up the possibility for a more
detailed Monte Carlo analysis. The potential issues with the signal detection would be the
identification of the emitted tau lepton through its decay products. Two promising tau decay
channels are the leptonic decay $\tau\to \mu \bar{\nu}_\mu\nu_\tau$ and the hadronic decay $\tau\to
\pi \mu \bar \nu_\tau$. In any case, the main background is expected to arise from the lepton flavor
conserving $\mu P\to \mu X$ process, whose signal could mimic that of the muon or the pion arising
from
the tau decay channels.

\acknowledgments{We would like to thank  SNI and Conacyt (M\'{e}xico) for financial
support. G. T. V. would like to thank VIEP-BUAP for support. The work of A. Bola\~nos is supported
by a postoctoral fellowship awarded by Conacyt.}

\end{document}